**Observation and Control of the Magnetic Photogalvanic Effect from Strongly Bound Excitons**

Xianfan Nie[1,2 †], Tarun Patel[1,2,3 †], Hyunggeun Lee[4], Chuangtang Wang[5], Meixin Cheng[1,6], Mingrui Lai[4], Bowen Yang[1,2], Matteo Pennacchietti[1,3], Shiyu Liu[7,8], Hechang Lei[7,8], Liuyan Zhao[5], Michael E. Reimer[1,3], Su Ying Quek[4], and Adam W. Tsen[1,6*]

[1]Institute for Quantum Computing, University of Waterloo, Waterloo, ON, Canada.
[2]Department of Physics and Astronomy, University of Waterloo, Waterloo, ON, Canada.
[3]Department of Electrical and Computer Engineering, University of Waterloo, Waterloo, Canada.
[4]Department of Physics, National University of Singapore, Singapore.
[5]Department of Physics, University of Michigan, Ann Arbor, MI, USA.
[6]Department of Chemistry, University of Waterloo, Waterloo, ON, Canada.
[7] School of Physics and Beijing Key Laboratory of Opto-electronic Functional Materials & Micro-nano Devices, Renmin University of China, Beijing 100872, China
[8] Key Laboratory of Quantum State Construction and Manipulation (Ministry of Education), Renmin University of China, Beijing, 100872, China
†These authors contributed equally
* Correspondence to: awtsen@uwaterloo.ca

**Abstract**

Photogalvanic effects arising from the quantum geometry of noncentrosymmetric materials are promising for next-generation light-harvesting devices that don't require a built-in electric field. Recent theories predict photogalvanic currents generated in magnetic systems with spin-dependent symmetry breaking as well as by bound exciton states, allowing for potential magnetic field control of the photoresponse and enhanced detection of deep sub-gap signals, respectively. We demonstrate the magnetic photogalvanic effect in a bilayer $CrI_3$ tunnel junction with both magnetic field switching and electric field tuning of interlayer symmetry. By controlling for the polarization and energy of light illumination, we disentangle the shift and injection current contributions and find that the peak response occurs under resonant excitation of strongly bound excitons in $CrI_3$. Our results can be captured within a many-body framework of the photogalvanic effect, while our devices function as tunable, multispectral helicity- and polarization-sensitive detectors that highlight the potential of 2D magnets for future optoelectronic applications.

## Main

The photogalvanic (or bulk photovoltaic) effect is a nonlinear photocurrent generation mechanism arising from the quantum geometry of electron bands[1]. Unlike the conventional photovoltaic effect, where photoexcited carriers must be separated by an electric field and transported to the contacts, photogalvanic currents can be generated directly during the optical transition, potentially enabling ultrafast and dissipationless devices[2,3]. The effect requires materials without an inversion center and originates from electronic states that acquire small positional shifts or momentum asymmetry upon optical excitation (giving rise to shift or injection currents, respectively). In the past, most studies have focused on free-carrier excitations in inherently noncentrosymmetric semiconductors and semimetals with above-gap illumination, which can be described within an independent particle framework of the underlying quantum geometry[2,4–13]. Recently, there has been much theoretical interest on two fronts: 1) photogalvanic effects in magnetic systems with spin-dependent inversion symmetry breaking, allowing for magnetic field control of nonlinear photocurrents[14–17], and 2) the possibility of enhanced photocurrent generation from bound exciton states under resonant excitation, a correlated many-body process that is not possible for ordinary photovoltaics without exciton dissociation[18–22].

There are few experimental works in either area. Photocurrent measurements performed on the magnetic topological insulator $MnBi_2Te_4$ do not find any contribution explicitly tied to the symmetries of the spin states[23]. Studies on nonmagnetic CdS and CuI report an enhanced shift current for one or more exciton states[24,25]; however, they all have relatively small binding energies ($\lesssim$ 60meV), making them more prone to dissociation and less useful for the transduction of deep sub-gap signals. In this work, we uncover a new magnetic photogalvanic effect originating from Wannier excitons in bilayer $CrI_3$[26], a system with magnetic-field-switchable symmetry states and extremely large exciton binding energy (reaching 1.1eV). Our tunnel junction geometry probes for the out-of-plane component of photocurrent and simultaneously provides fine-tuning of the layer parity using electric fields. Controlling the light polarization further allows us to isolate the shift and injection current contributions to the magnetic photogalvanic effect and demonstrate, in distinction with existing theory, that both can arise from strongly bound excitons. Our work establishes 2D spin systems as a unique platform for magnetically and electrically tunable optoelectronic devices with polarization sensitivity.

The symmetry of bilayer $CrI_3$ in the magnetic state is summarized in Fig. 1a. The ground state exhibits antiferromagnetic (AFM) layer coupling, producing two spin configurations that are usually degenerate in measurement. Each AFM state breaks parity ($\mathcal{P}$) or inversion symmetry and time reversal ($\mathcal{T}$) symmetry separately but preserves the combined $\mathcal{PT}$ symmetry. In contrast, the spin-polarized FM states retain $\mathcal{P}$ but break $\mathcal{T}$ and $\mathcal{PT}$ symmetry. Importantly, this implies that the magnetic photogalvanic effect should only be observed for the AFM states. Furthermore, as the two AFM configurations are time-reversal partners ($\mathcal{T}$AFM1 = AFM2), switching between them can reverse the photocurrent for time-reversal-odd nonlinear processes[27]. A magnetic field thus provides a ternary knob to engage, reverse, and switch off this photogalvanic effect.

Such a field-switchable photogalvanic effect has been previously predicted in bilayer $CrI_3$ in a lateral device geometry[14], but never experimentally observed, likely due to the system's highly insulating nature below the magnetic transition temperature[28]. Figure 1b shows our alternative device and measurement geometry incorporating bilayer (2L) $CrI_3$ as a magnetic tunnel barrier between graphene electrodes with hexagonal boron nitride (hBN) encapsulation (see Methods for fabrication details). The tunneling magnetoresistance has been shown to be highly sensitive to the interlayer coupling (AFM or FM)[28–31], while (linear) photocurrent effects in few-layer $CrI_3$ junctions have also been shown to be dependent on the spin configuration[32,33]; however, nonlinear photogalvanic currents have not been reported and likely are only clearly seen in the bilayer (as we shall discuss).

The differential reflectance of 2L-$CrI_3$ is shown in Fig. 1c. The peaks between 1.5-2.0eV are attributed to bright Wannier-type excitons, while the continuum states are above 2.6eV[26,34,35]. The A, B+, and B- excitons are thus tightly bound with binding energies of 1.1, 0.8, and 0.7eV, respectively. For most of this work, we excite our devices with a continuous 633nm (1.96eV) laser close to the B- exciton resonance at 2K (see Methods for experimental details). We also report the spectral dependence of the photogalvanic effect in the range of 1.3-2.2eV, which will be described later on. The main panel of Fig. 1d shows the current-voltage characteristics of Device 1 with and without laser illumination, while the inset explicitly plots the photocurrent $I_{ph} = I_{light} - I_{dark}$, which is generally mix of linear and nonlinear contributions. We see a sizable photocurrent at zero bias and an applied bias of ~-10mV is needed to null the photocurrent (see Extended Data Fig. 1 for laser power dependence). Figure 1e shows a scanning zero-bias photocurrent image of Device 2, where $I_{ph}$ is mostly confined to illumination of the junction, as expected.

The linear and nonlinear photocurrents can be disentangled by controlling both the light polarization and spin states. As our device geometry probes for out-of-plane photocurrent, the magnetic symmetries of bilayer $CrI_3$ allow for effectively only two relevant terms under normal incidence: $j_z = 2\chi_{zxy}E_xE_y^* + 2\chi_{zxy}^*E_yE_x^*$ (see Supplementary Information for full expression)[36,37]. For materials with $\mathcal{PT}$ symmetry, these terms originate from two microscopic processes termed the magnetic injection current (MIC) and magnetic shift current (MSC): $\chi = Re(\chi) + iIm(\chi) = \tau\eta^{MIC} + \sigma^{MSC}$ [15,16]. Both are time-reversal odd; however, for the free carrier case, the former scales with scattering time $\tau$ and the latter is scattering-independent. For bound excitons, $\tau$ should instead be interpreted as the effective exciton lifetime.

Under left- or right-circularly polarized (LCP or RCP) excitation with $E_x = E$, and $E_y = \pm iE$, only the MSC contributes: $j_z^{LCP/RCP} = \mp 4i\sigma_{zxy}^{MSC}E^2$. For a given AFM order, we see that the nonlinear photocurrent changes sign when switching the light handedness, while for fixed polarization, current also changes sign when switching between the two AFM states, as $\mathcal{T}\sigma_{zxy}^{MSC}(\mathrm{AFM1}) = \sigma_{zxy}^{MSC}(\mathrm{AFM2}) = -\sigma_{zxy}^{MSC}(\mathrm{AFM1})$. These are two hallmark signatures for the circular magnetic photogalvanic effect. To demonstrate this, we first confirm our device operation under an out-of-plane magnetic field. Figure 2a and 2b show the tunneling magnetoresistance and reflection magnetic circular dichroism (RMCD), which measures the relative interlayer spin coupling and net out-of-plane magnetization, respectively, while sweeping the magnetic field back and forth to access the four spin configurations in Device 1. In both cases, the two AFM states around zero magnetic field with symmetric tunneling barriers and quenched magnetization have nearly degenerate values. This symmetry is lifted in the circular photogalvanic measurements shown in the upper panels of Fig. 2c. For LCP light, there is a sizeable photocurrent gap between the two AFM states accessed between field sweep up (red trace above) and down (blue trace below). This gap is reversed (blue trace above, red trace below) for RCP light. These features are in accordance with the symmetry argument for nonlinear photocurrents (summarized schematically in the insets); although, they sit on top of a linear photocurrent background that is also spin-dependent[32].

To isolate the nonlinear contribution, we first plot the photocurrent MCD ($I_{ph}^{MCD} = I_{ph}^{LCP} - I_{ph}^{RCP}$) in analogy with RMCD in the lower panel of Fig. 2c. This reveals clear plateaus for the four spin states and nearly opposite photocurrent values for the time-reversal partners. The large gap seen for the AFM states is consistently observed across several bilayer devices (see Extended Data Fig. 2) and closes above the magnetic transition temperature (see Extended Data Fig. 3), indicating that it is a general feature of spin-dependent inversion symmetry breaking of the AFM phase. It is interesting to note that this photogalvanic feature appears to be unique to 2L-$CrI_3$. Trilayer $CrI_3$ restores inversion symmetry and so only yields linear photocurrent behavior[32], while four-layer $CrI_3$ exhibits no substantial photocurrent gap between its AFM states (see Extended Data Fig. 4).

Unlike the AFM states, the two FM states in 2L-$CrI_3$ are, in principle, inversion-symmetric and so are not expected to contribute nonlinear photogalvanic behavior. However, any slight asymmetry between the two graphene electrodes could lead to a built-in electric field that breaks parity between the two $CrI_3$ layers. As the FM states possess circular dichroism (as shown in RMCD), the field can also yield $I_{ph}^{MCD}$ through the ordinary photovoltaic effect. Consistent with this, we see in Extended Data Fig. 2 that the photocurrent difference for the two FM states varies strongly from sample to sample, whereas the AFM gap is more robust.

To control for this built-in field, we show photocurrent MCD while applying several different biases to Device 3 (+10mV, 0, and -10mV) in Fig. 3a. We see that while the $I_{ph}^{MCD}$ difference between the AFM states ($\Delta_{AFM}$) is relatively robust, the difference for the FM states ($\Delta_{FM}$) changes strongly and nearly vanishes for -10mV. In Fig. 3b, we further plot these quantities continuously as a function of applied bias. Indeed, we find that both $\Delta_{AFM}$ and $\Delta_{FM}$ can be fine-tuned through an out-of-plane electric field, although the two show opposite trends. $\Delta_{FM}$ is strongly enhanced with increasing positive bias and zeroed with negative bias (near -10mV). In contrast, $\Delta_{AFM}$ is larger for increasing negative bias and decreases for positive bias, but it cannot be shut off entirely within our voltage range. The electric field thus offers continuous tunability over both linear and nonlinear photocurrents in our device geometry.

We now discuss observation and control of the linear (polarization) magnetic photogalvanic effect in 2L-$CrI_3$. Under linear polarized light with $E_x = Ecos\theta$ and $E_y = Esin\theta$ (where $x$ is defined to be the $C_2$ rotation axis and $y$ is defined to be along the mirror plane—see schematic inset in Fig. 3c), only the MIC contributes to the nonlinear photocurrent: $j_z = 4\tau\eta_{zxy}^{MIC}E^2 sin\theta cos\theta$. For each AFM state, we thus expect two full-period oscillations as the polarization angle is swept a full circle. Furthermore, as $\mathcal{T}\eta_{zxy}^{MIC}(\text{AFM1}) = \eta_{zxy}^{MIC}(\text{AFM2}) = -\eta_{zxy}^{MIC}(\text{AFM1})$, we also expect the oscillations for the two AFM states to be fully out of phase for perfectly balanced layers. To demonstrate these hallmark features, in the main panels of Fig. 3c we plot $I_{ph}$ for Device 3 as a function of the linear light polarization angle relative to the $x$ ($C_2$ rotation) axis for both AFM states and under three applied biases (+10mV, 0, and -10mV). The in-plane crystalline axes were identified using optical second harmonic generation measurements (see Methods and Extended Data Fig. 5). In all cases, we observe oscillations with two complete periods; however, the phases of the oscillations and background levels generally change with bias. We attribute the former to layer (a)symmetry and latter to the ordinary photovoltaic effect, both of which are affected by the applied electric field. In particular, for -10mV (the same voltage where $\Delta_{FM}$ is nearly zero), the background is suppressed and oscillations for the two AFM states are almost exactly out of phase, as expected from our symmetry analysis.

We have repeated these polarization-angle-dependent measurements on the two AFM states for many more biases, and the complete results are plotted in Extended Data Fig. 6. For each data set, we have fit the angle dependence to the form $I_{ph} = Asin(\theta + \phi)\cos(\theta + \phi)$. The amplitudes $A$ and phase shifts $\phi$ for both AFM states are plotted in Fig. 3d as a function of bias. We see that $A$ remains relatively constant and similar for AFM1 and AFM2, while $\phi$ can be widely tuned. Around -10mV, the two phases are close to the ideal conditions of 0 and 90 degrees; and with increasing positive bias, the phase difference between the two states $\Delta\phi = \phi_{\mathrm{AFM1}} - \phi_{\mathrm{AFM2}}$ decreases.

Taken together, our voltage-dependent photocurrent measurements not only demonstrate electrical control of both circular and linear magnetic photogalvanic effects in bilayer $CrI_3$ but further confirms the bound exciton mechanism. If exciton dissociation was necessary to generate the nonlinear photocurrents, we would expect these components to be suppressed with negative bias together with the background photocurrent. Instead, the amplitude for the linear photogalvanic contribution remains constant, while $\Delta_{AFM}$ for the circular contribution increases at negative bias. These results indicate that, in contrast to the ordinary photovoltaic effect which requires exciton dissociation by an electric field, the nonlinear photogalvanic effects are generated from the bound exciton state itself and do not require any built-in or applied electric field.

We now explicitly show our photocurrent spectral dependence and theoretically analyze the role of strongly bound excitons in bilayer $CrI_3$ to manifest the nonlinear photogalvanic response. In Fig. 4a, we show zero-bias photocurrent MCD (normalized to incident light power) vs excitation energy around the three excitonic transitions for the two AFM states. Here, we can gauge the strength of the circular magnetic photogalvanic effect (MSC contribution) from the gap between the two spectra. The gaps are largest at the exciton transitions and become negligible away from these resonances, indicating that the MSC in this energy regime is mainly generated by bound exciton states. From the peak photocurrent values, we extract a magnetic shift conductivity of $\left|\sigma_{zxy}^{MSC}\right| \sim 3.7 \times 10^{-8}, 1.5 \times 10^{-7}$, and $1.8 \times 10^{-7} \frac{A}{V^2}$ for the A, B+, and B- excitons, respectively. By comparison, the value for the B- exciton is about an order of magnitude larger than the shift conductivity reported for thin film $BaTiO_3$[38], a canonical photogalvanic material with pure structural symmetry breaking. In Fig. 4b, we similarly plot the spectral dependence for linearly polarized light (at three important polarization angles). The overall background can be attributed to the ordinary photovoltaic effect, while the difference between the two AFM states is due to the linear (polarization) magnetogalvanic effect (MIC contribution). At -45 degrees, we see the largest difference again occurs at the exciton resonances, indicating that MIC is also dominated by bound excitons. The difference nearly vanishes at 0 degrees for all light energies and flips sign at +45 degrees, as expected from symmetry. At the exciton resonances, we extract a magnetic injection conductivity of $\left|\tau\eta_{zxy}^{MIC}\right| \sim 3.8 \times 10^{-8}, 4.8 \times 10^{-8}$, and $7.5 \times 10^{-8} \frac{A}{V^2}$ for the A, B+, and B- excitons, respectively.

The possibility of shift currents generated by bound excitons has been predicted theoretically in the literature[18–22]; however, to our knowledge, there have not been any theoretical or experimental reports of exciton-driven injection currents. Here, we provide an intuitive understanding of both effects, while the full expressions for $\sigma_{zxy}^{MSC}$ and $\eta_{zxy}^{MIC}$ can be found in the Supplementary Information. The exciton can be considered as a two-body state involving the excited electron and corresponding hole. It has recently been shown that the shift current originates from a many-body shift vector arising from the quantum geometry of the excitonic state, which can be interpreted as the expectation value of positional difference between the excited electron and the hole, i.e. $\langle r_e - r_h \rangle$[21]. When this value is nonzero, there is a small shift of charge relative to the ground state, which manifests as current for a large number of excitons generated by the laser field. This is formally analogous to, but distinct from the shift vector within the independent particle approximation (IPA). where the nonlinear response is built from independent valence-to-conduction transitions at each crystal momentum, and the shift currents are governed by single-particle positional shifts[4,5,39]. In the many-body exciton picture, by contrast, the excitonic state is a coherent superposition of correlated electron–hole pairs over momentum space. Electron–hole interaction therefore changes not only the transition energy and oscillator strength, but also the shift vector. Similarly, within the IPA, the injection current for free carriers can be expressed in terms of the difference in velocities of the conduction and valence band Bloch states[39,40]. For excitons, it can be shown that the injection current arises from $\langle v_e - v_h \rangle$, the expectation value of the relative velocity of the electron and hole within the bound exciton. Within the exciton lifetime $\tau$, this also leads to a net movement of charge. Thus, both shift and injection currents can be generated for either free carriers or bound excitonic states without the need for their dissociation (see schematics in Fig. 4c for the excitonic case). Using the absorbance at the excitonic transitions of bilayer $CrI_3$[34], together with the measured values of $\sigma_{zxy}^{MSC}$ and $\eta_{zxy}^{MIC}$, we estimate the effective out-of-plane many body shift vector and relative out-of-plane velocity of the electron and hole for the A, $B^+$, and $B^-$ excitons. Assuming isotropic matrix elements, we obtain $|\langle r_e - r_h \rangle| \sim 6 \times 10^{-4}, 4 \times 10^{-2}, 7 \times 10^{-3}$ Å and $|\langle v_e - v_h \rangle| \sim 4 \times 10^{-2}, 2 \times 10^{0}, 4 \times 10^{-1}$ m/s, respectively.

In conclusion, we have discovered shift and injection photogalvanic currents in a 2D magnet with behavior that can be controlled by magnetic and electric fields. The effects are enhanced at the resonances of strongly bound exciton states and do not require exciton dissociation into free carriers, in accordance with recent theoies that consider the many-body interaction. Our work catalyzes future theoretical efforts on the study of correlated (magnetic) photogalvanic effects. Our results also pave the way for dissipationless and ultrafast photogalvanic devices based on 2D magnets with tunable photon helicity and polarization sensitivity.

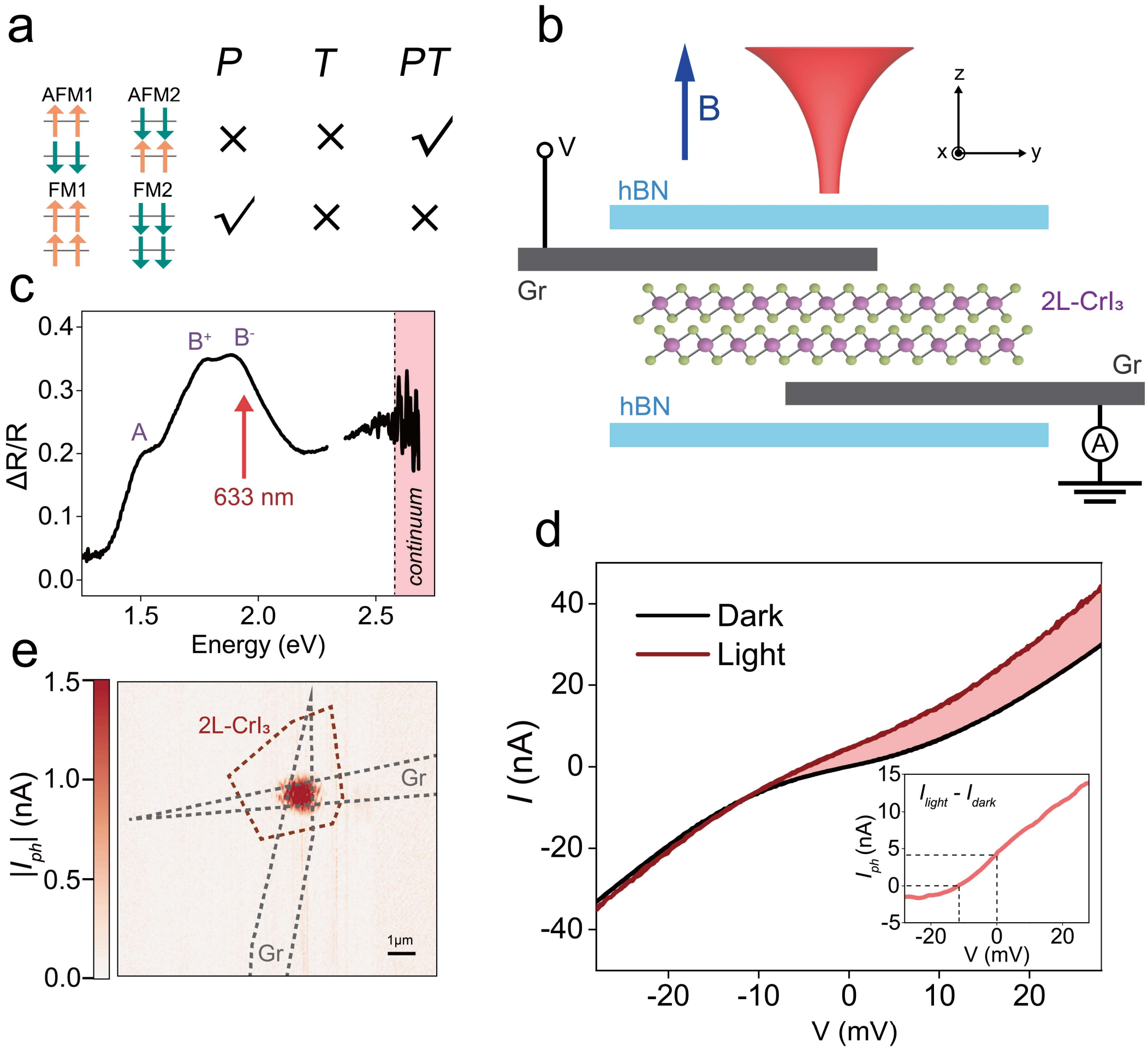

**Figure 1 | Device geometry of bilayer CrI₃ tunnel junctions and photocurrent measurement scheme.**

**a** The four possible magnetic states in 2L-$CrI_3$ and their symmetries. **b** Sideview schematic of 2L-$CrI_3$ vertical junction device with graphene (Gr) contacts and hexagonal boron nitride (hBN) encapsulation. Measurement configuration of out-of-plane photocurrent under normal laser incidence and out-of-plane magnetic field. **c** Differential reflectance spectrum of 2L-$CrI_3$, highlighting resonances from strongly bound Wannier-type excitons (A, $B^+$, $B^-$) below the continuum bandgap (2.6eV). 633nm laser wavelength excitation energy is indicated with a red arrow. **d** Current–voltage characteristics of a representative device (Device 1) under dark conditions and under laser illumination. Inset shows the photocurrent $I_{\text{ph}} = I_{\text{light}} - I_{\text{dark}}$ as a function of bias. The zero-bias photocurrent and nulled photocurrent bias are indicated. **e** Spatial mapping of the zero-bias photocurrent for Device 2 obtained by scanning laser microscopy.

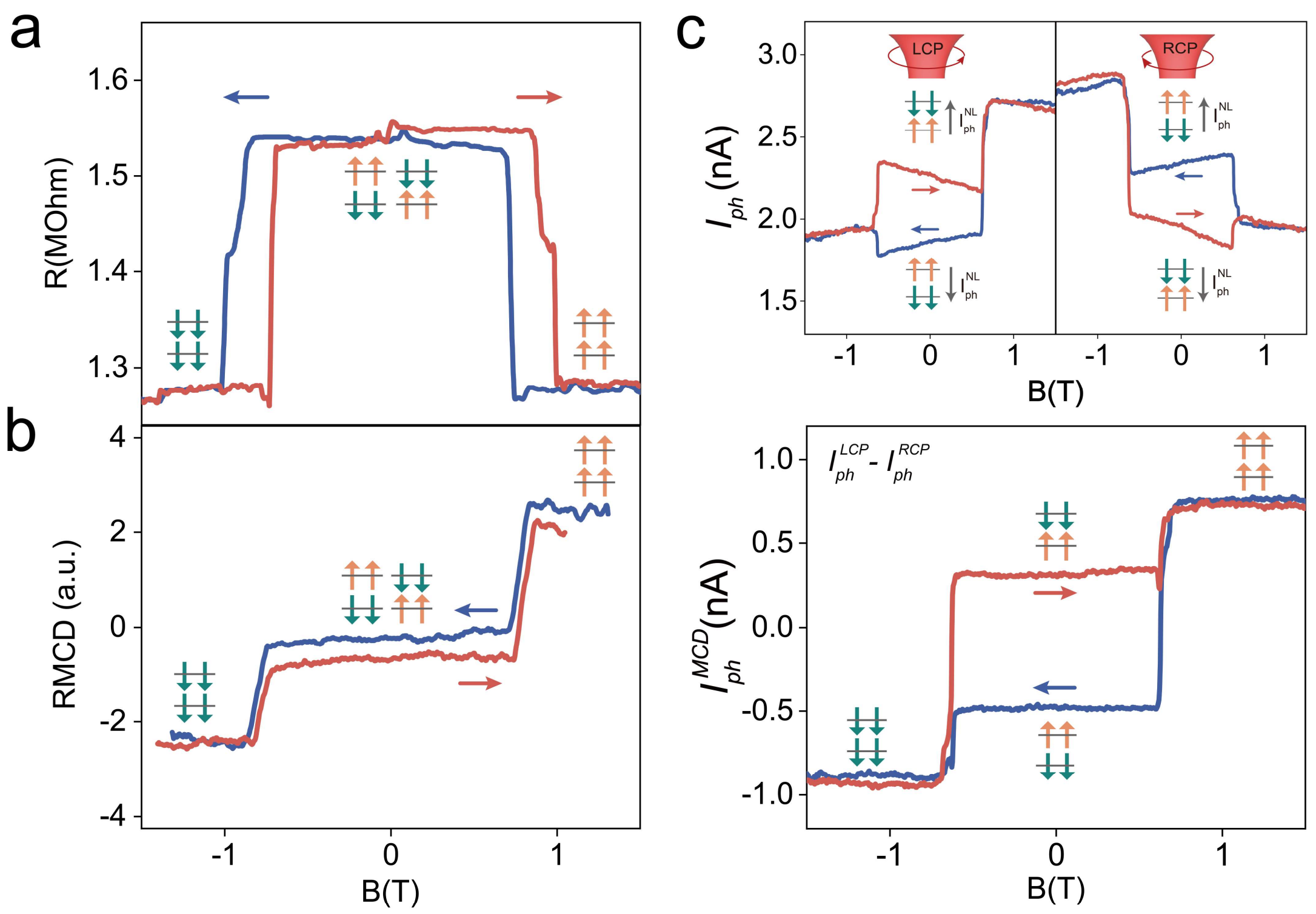

**Figure 2 | Magnetic-field-switching of spin states and circular magnetic photogalvanic effect in bilayer $CrI_3$.**

**a** Tunneling magnetoresistance of 2L-$CrI_3$ Device 1 as a function of out-of-plane magnetic field *B*. The two FM states exhibit lower tunneling resistance compared with the AFM states. **b** Reflection magnetic circular dichroism (RMCD) versus *B* for the same device. This measurement probes for the net out-of-plane magnetization of the sample. **c** Top: Device 1 photocurrent $I_{\mathrm{ph}}$ under left- and right-circularly polarized light excitation (LCP/RCP) versus *B*. Schematics illustrate the expected flipping of nonlinear photocurrent signals with changing light helicity and AFM state. Bottom: photocurrent MCD $I_{\mathrm{ph}}^{\mathrm{MCD}} = I_{\mathrm{ph}}^{\mathrm{LCP}} - I_{\mathrm{ph}}^{\mathrm{RCP}}$ as a function of *B* defined in analogy with RMCD. The two AFM states show substantial and opposite plateau signals in photocurrent MCD, but similar quenched magnetization in RMCD.

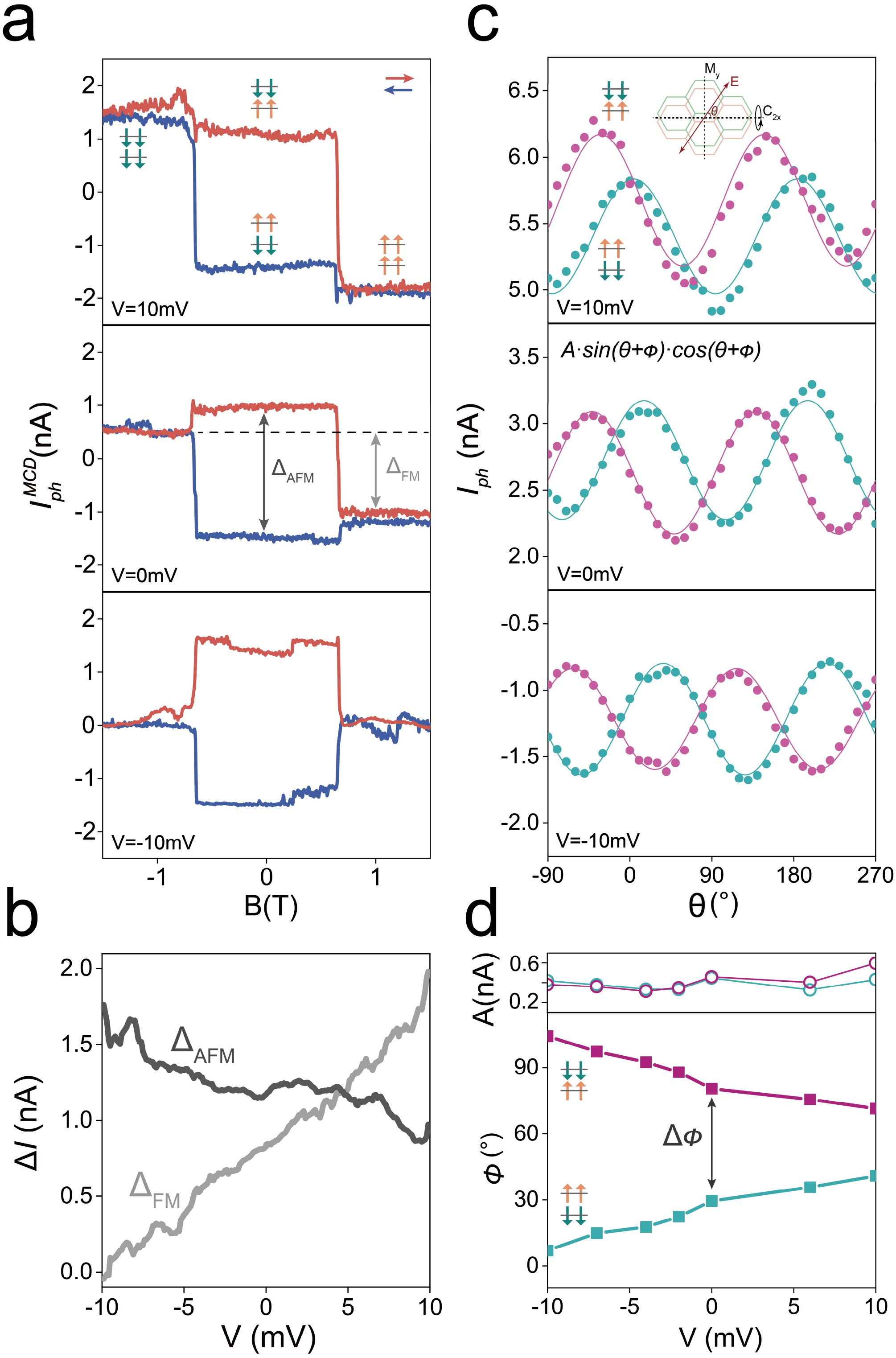
a
b
c
d
V=10mV
V=0mV
V=-10mV
B(T)
$I_{ph}^{MCD}$(nA)
$\Delta_{AFM}$
$\Delta_{FM}$
ΔI (nA)
V (mV)
$I_{ph}$ (nA)
θ(°)
A·sin(θ+Φ)·cos(θ+Φ)
A(nA)
Φ (°)
ΔΦ
$M_y$
E
θ
$C_{2x}$

**Figure 3 | Electrical control of circular and linear magnetic photogalvanic effects in bilayer $CrI_3$.**

**a** Photocurrent MCD field sweeps measured for three different biases *V* = +10, 0, -10mV applied across the junction (Device 3). The difference between the FM states ($\Delta_{\mathrm{FM}}$) can be widely tuned, while that for the AFM states ($\Delta_{\mathrm{AFM}}$) remains relatively robust. **b** $\Delta_{\mathrm{FM}}$ and $\Delta_{\mathrm{AFM}}$ vs continuous bias extracted from photocurrent field sweeps showing opposite trends. $\Delta_{\mathrm{FM}}$ strongly decreases with decreasing (negative) bias and is zeroed near -10mV. $\Delta_{\mathrm{AFM}}$ slightly decreases with increasing (positive) bias but cannot be quenched within the voltage range. **c** Linear-polarization dependence of $I_{\mathrm{ph}}$for the two AFM states at *V* = +10, 0, -10mV for the same device. The polarization angle $\theta$ is defined relative to the crystallographic $C_2$ axis (schematic inset). Solid curves are fits to $A\sin(\theta+\phi)\cos(\theta+\phi)$. The overall phase $\phi$ for the two AFM states shift with bias, becoming almost completely out of phase for *V* = -10mV. **d** Fitted parameters $A$ and $\phi$ for the two AFM states vs *V*. The amplitudes A for both AFM states remain relatively constant, while their phases $\phi$ shift in opposite directions ($\Delta\phi$ decreases) with increasing (positive) bias.

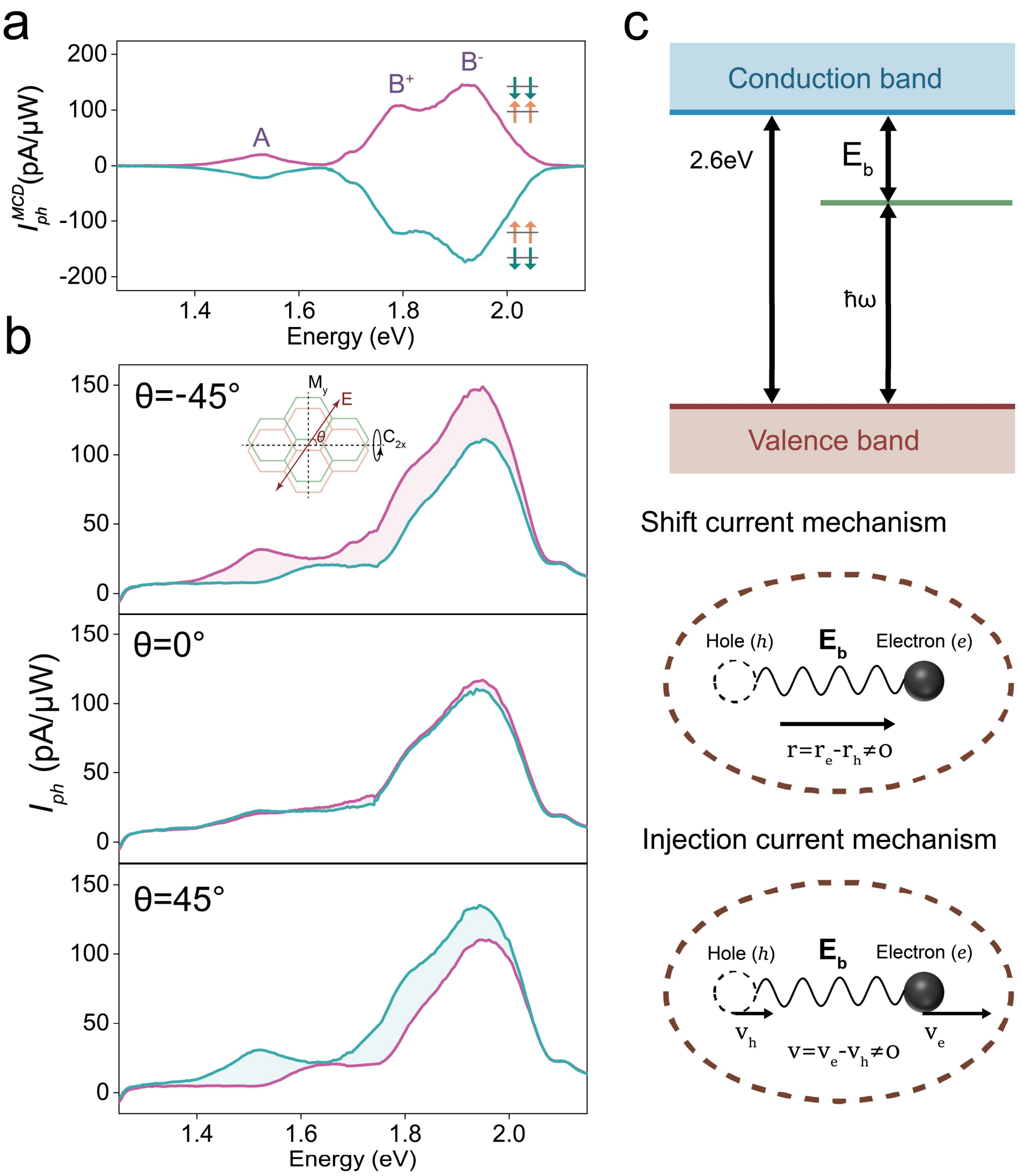

**Figure 4 | Spectral dependence of circular and linear magnetic photogalvanic responses and schematic illustrations of the shift and injection current mechanisms for bound excitons.**

**a** Spectral dependence of the zero-bias photocurrent MCD signal $I_{\mathrm{ph}}^{\mathrm{MCD}}$(normalized by incident power) for the two AFM states. The difference is largest at the energies of the strongly bound exciton states and disappear away from resonance. **b** Similar spectral dependence for linearly polarized light at three polarization angles ($\theta = -45°, 0°, +45°$). In accordance with symmetry, the two AFM spectra nearly overlap at $\theta = 0°$, while their differences flip sign between $\theta = \pm 45°$. The differences are again largest at the bound exciton resonances. **c.** Conceptual illustration of excitonic shift and injection current mechanisms. $E_{\mathrm{b}}$ and $\hbar\omega$ represent the exciton binding energy and laser excitation energy, respectively. $\mathbf{r}_{e(h)}$ and $\mathbf{v}_{e(h)}$ denote the position and velocity of the electron (hole).

## Methods

### Crystal synthesis

$CrI_3$ single crystals were grown by the chemical vapor transport method. $CrI_3$ polycrystals were placed inside a silica tube with 200mm length and 14mm inner diameter. After evacuation to 0.01Pa and sealing, the tubes were moved into a two-zone horizontal furnace. The temperature of the source (growth) zone was slowly raised to 873–993K (723–823 K) over a 24hr period and then held there for 150hrs.

### Device fabrication

$CrI_3$, graphene/graphite (HQ Graphene), and h- BN (HQ Graphene) flakes were exfoliated onto silicon dioxide substrates inside a nitrogen-filled glove box ($P_{\mathrm{O_2}}, P_{\mathrm{H_2O}} < 0.1$ppm). Metal contacts (17nm Au/3nm Ti) and wire-bonding pads (40nm Au/5nm Ti) were pre-fabricated on sapphire by photolithography and electron beam deposition. The device heterostructures were assembled inside the glove box by sequential dry transfer using a polycarbonate coated polydimethylsiloxane stamp. The junction area was kept small (< 1$\mu\mathrm{m}^2$) to prevent electrical breakdown.

### Photocurrent measurements

Scanning photocurrent measurements were conducted in a Montana C2 Cryostation at 8K. A 632.8nm wavelength HeNe laser was focused to a diffraction limited spot (~1 μm) using an objective lens with numerical aperture (NA) of 0.55. The beam was raster-scanned across the device using a pair of galvanometric mirrors in combination with a telescopic relay lens system. The photocurrent was recorded with an NF Corporation CA5351 current preamplifier, while the reflected signal was collected simultaneously with a silicon photodiode.

Magnetic-field-dependent photocurrent measurements were conducted in an attocube attoDRY 2100 cryostat at 1.6K. For polarization-dependent measurements, a 632.8nm HeNe laser was mechanically chopped at ~1kHz and focused to a diffraction-limited spot (~1μm) using an objective with NA = 0.81. The incident beam was first linearly polarized using a Glan–Laser polarizer. Subsequently, the polarization axis was either rotated using a half-wave plate, or the light was converted to circular polarization with a quarter-wave plate. The laser spot was positioned on the device using a white-light imaging system together with an $x$–$y$–$z$ piezoelectric stage, and the photocurrent was detected with a Stanford Research

Systems SR865 lock-in amplifier referenced to the chopping frequency. When applicable, a d.c. bias was applied across the two graphene electrodes using a Keithley 2450 source measure unit. For continuous bias-dependent measurements, the resulting current was recorded directly using the Keithley source measure unit. For photocurrent spectroscopy, the excitation source was replaced by an NKT supercontinuum laser combined with a tunable bandpass filter, while the same optical setup and photocurrent detection scheme were maintained, and the excitation wavelength was swept over the desired range.

**Reflection magnetic circular dichroism (RMCD) measurements**

RMCD measurements were performed in the same attoDRY 2100 cryostat at 1.6K. A 632.8nm HeNe laser was mechanically chopped at ~1kHz and focused to a diffraction-limited spot (~1μm) using an NA = 0.81 objective. The incident polarization was modulated between left- and right-handed circular states with a photo-elastic modulator (Hinds Instruments PEM-200) operated at 50kHz. The reflected signal was collected through the same objective and detected with a photodiode and a Stanford Research Systems SR865 lock-in amplifier referenced to 50kHz.

**Differential reflectance spectroscopy**

Differential reflectance spectroscopy was performed in a Montana C2 Cryostation at 8K using a Bruker IFS 66v/S FTIR spectrometer coupled with a YSL supercontinuum laser. Illumination from the supercontinuum laser was sent through the FTIR beamsplitter and then focused onto the bilayer $CrI_3$ sample by a NA=0.5 reflective objective. The reflected signal was collected and detected by a silicon photodiode. The differential reflectance spectrum was calculated as $\Delta R/R = (R_{\mathrm{sample}} - R_{\mathrm{ref}})/R_{\mathrm{ref}}$, where $R_{\mathrm{ref}}$ is the reflectance from a nearby hBN-covered reference region.

**Tunneling magnetoresistance measurements**

Tunnelling resistance was measured in the same attoDRY 2100 cryostat at 1.6K using an SR865 lock-in amplifier.

**Rotational-anisotropy second harmonic generation (RA SHG) measurements**

RA SHG is carried out by rotating the polarization of linearly polarized incident light and analyzing its corresponding co-polarization SHG intensity (parallel channel). The incident light is 800nm wavelength from an 80MHz laser oscillator with a pulse duration of 100fs, which is focused onto the bilayer $CrI_3$ device with a beam size around 1μm. The reflected 400nm SHG light is directed to a photomultiplier tube for SHG counts measurement. The resulted RA SHG was fitted using the simulated functional form of electric-dipole SHG under the *2/m'* point group: $I^{2\omega}_{parallel}(\varphi) = \left[\chi^{ED}_{xxx} cos^3(\varphi) + \left(2\chi^{ED}_{yxy} + \chi^{ED}_{xyy}\right) sin^2(\varphi)\cos(\varphi)\right]^2$,where $I^{2\omega}_{parallel}(\varphi)$ is the SHG intensity in the parallel channel, $\varphi$ is the azimuthal angle of the polarization with respective to the $C_{2x}$ axis, and $\chi^{ED}_{xxx}$ , $\chi^{ED}_{yxy}$ and $\chi^{ED}_{xyy}$ are the nonzero susceptibility tensors under the normal incidence geometry.

**Acknowledgements**

We thank Prof. Li Yang, Prof. John Sipe, and Prof. Wencan Jin for helpful discussions. AWT acknowledges support from the US Air Force Office of Scientific Research (FA9550-24-1-0360), the Natural Sciences and Engineering Research Council of Canada (NSERC), and the Transformative Quantum Technologies Program. This work is partially supported by NUS and the National Research Foundation (NRF), Singapore, under the NRF medium-sized centre programme, as well as by A*STAR, Singapore under Project ID M24N7c0095. Calculations were performed on the National Supercomputing Centre, Singapore. MER acknowledges support from NSERC Discovery, NSERC Quantum Alliance and Ontario Research Fund - Research Excellence. LZ acknowledges support from the U.S. Department of Energy, Office of Basic Energy Sciences, under Award DE-SC0024145. HCL acknowledges support from the National Key R&D Program of China (Grants No. 2023YFA1406500 and No. 2022YFA1403800) and the National Natural Science Foundation of China (Grant No. 12274459).

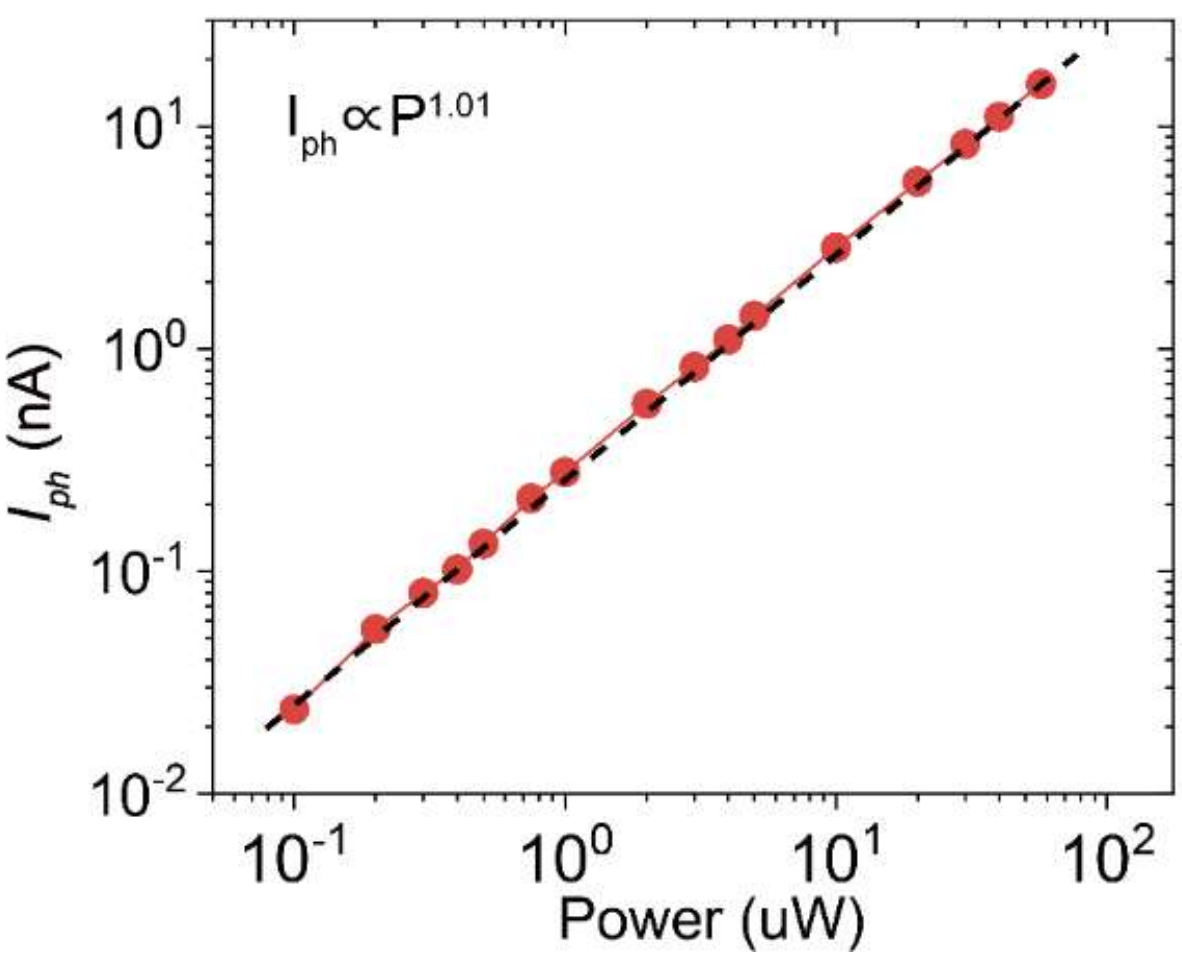

**Extended Data Fig. 1 | power dependence of zero-bias photocurrent.** Zero-bias photocurrent $I_{ph} = I_{light} - I_{dark}$ of Device 1 as a function of incident laser power in log-log cale. The data is well-described by a power-law fit, $I_{ph} \propto \mathrm{P}^{1.01}$, indicating a linear dependence on excitation power over three orders of magnitude. For all photocurrent measurements with the HeNe laser at fixed power, we have kept the excitation below $50\mu\mathrm{W}$.

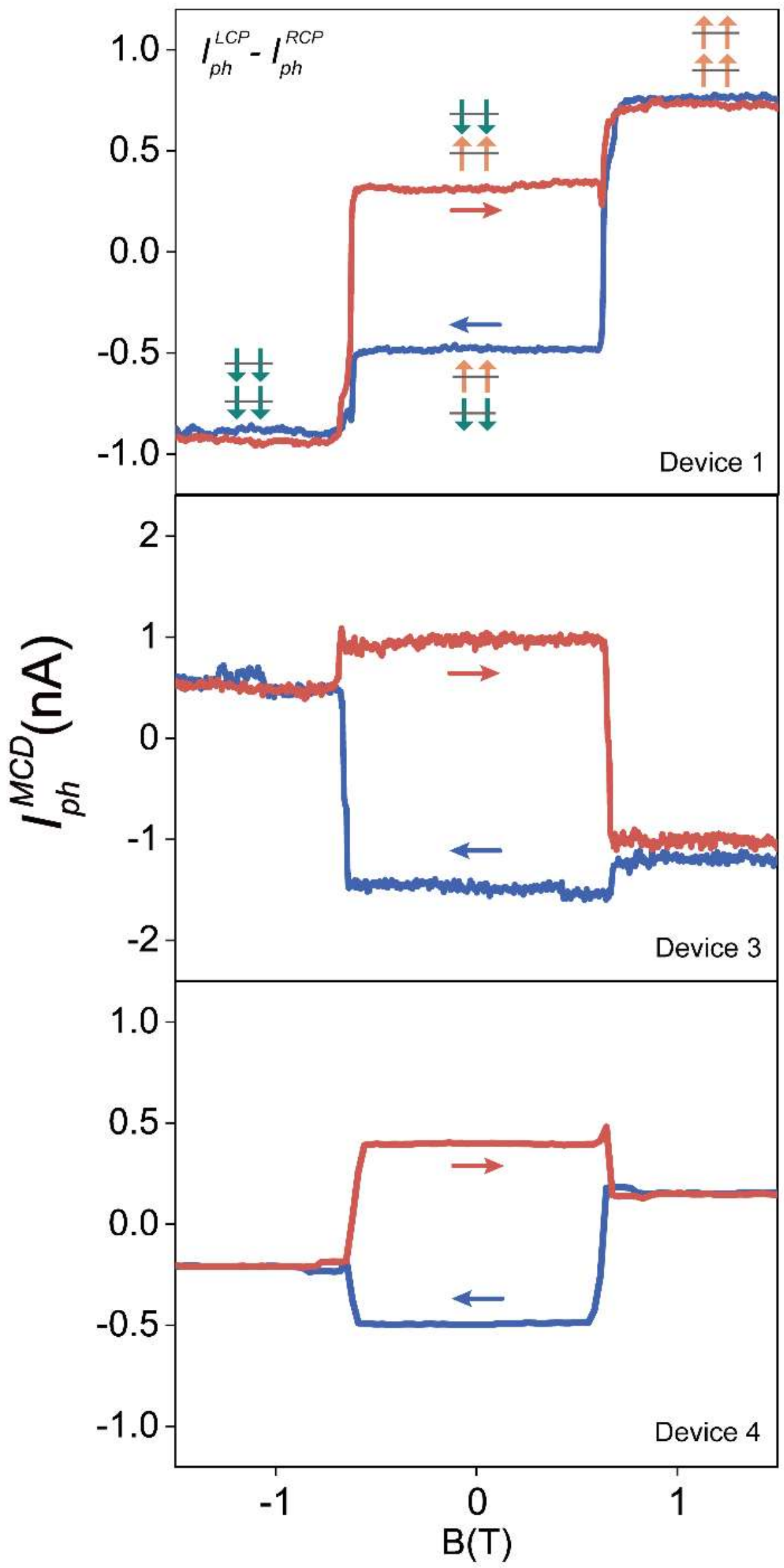

**Extended Data Fig. 2 | Circular magnetic photogalvanic response reproduced across multiple bilayer $CrI_3$ devices.** Photocurrent MCD ($I_{ph}^{MCD} = \mathrm{I}_{ph}^{LCP} - \mathrm{I}_{ph}^{RCP}$) as a function of out-of-plane magnetic field for Devices 1, 3, and 4. All three bilayer devices show a substantial splitting between the two AFM states.

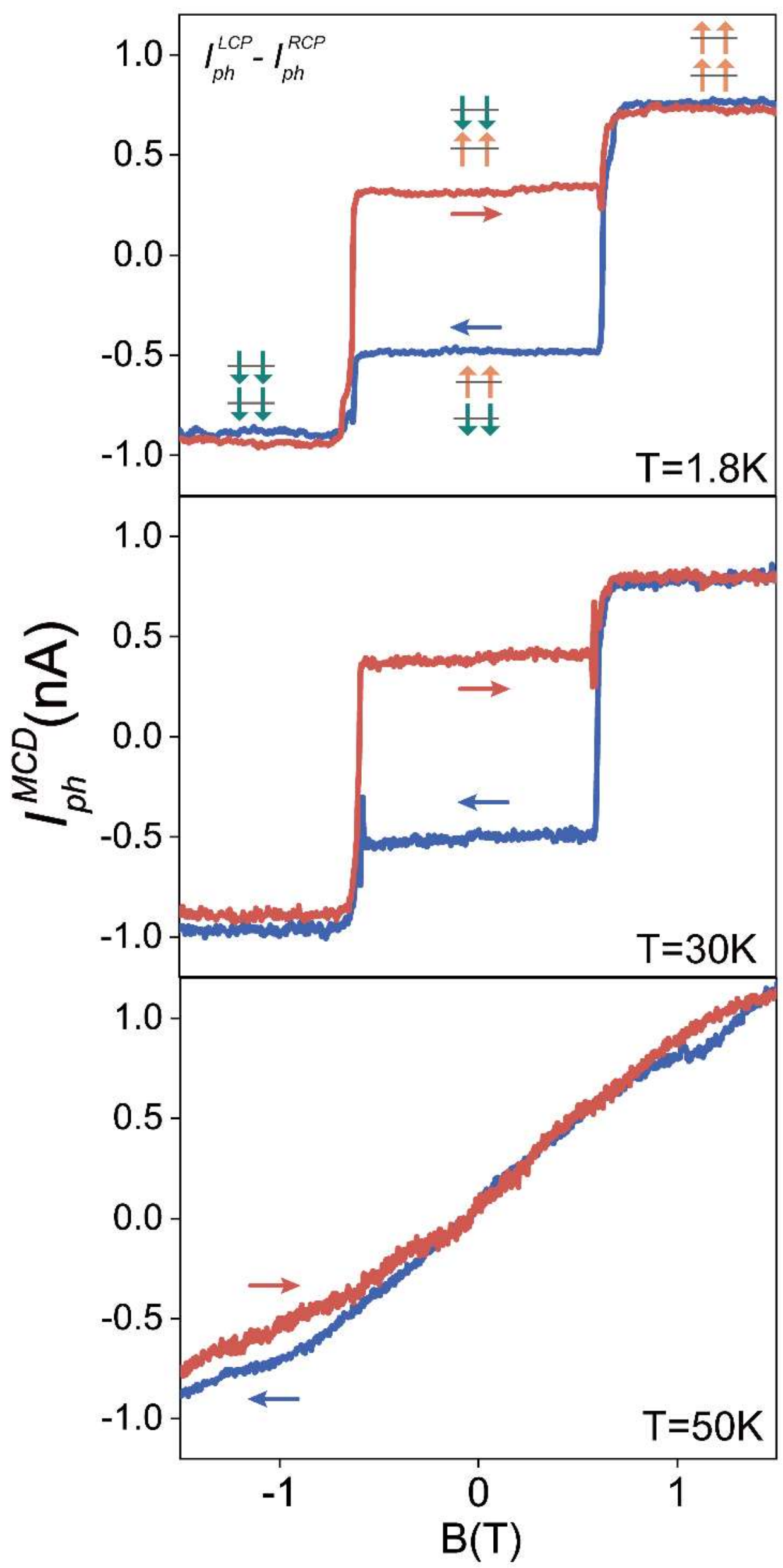

**Extended Data Fig. 3 | Temperature dependence of the circular magnetic photogalvanic effect.** Photocurrent MCD $I_{ph}^{MCD}$ measured on Device 1 at 1.8, 30, and 50K. The splitting between the two AFM states is clearly resolved at 1.8K and 30K, and disappears at 50K, consistent with the loss of spontaneous magnetic order above the transition temperature (45K)[41].

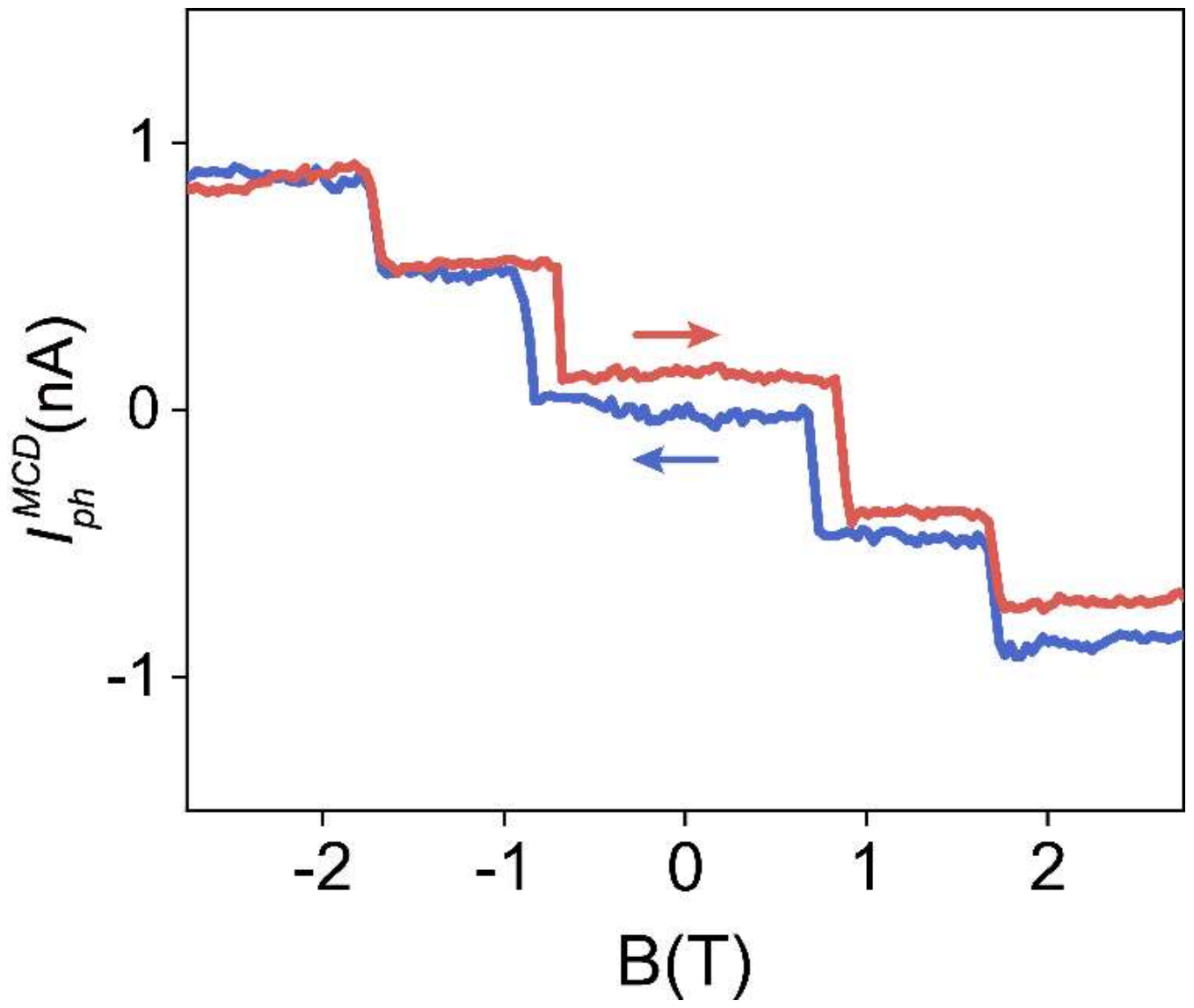

**Extended Data Fig. 4 | Absence of photocurrent MCD splitting for AFM states in a four-layer $CrI_3$ device.** Photocurrent MCD as a function of out-of-plane magnetic field measured on a four-layer $CrI_3$ tunneling device. Although multiple magnetic transitions are resolved in the field sweep, no substantial photocurrent gap is observed between the AFM states, in contrast to the bilayer devices. This indicates that the magnetic photogalvanic response is most visible in bilayer $CrI_3$.

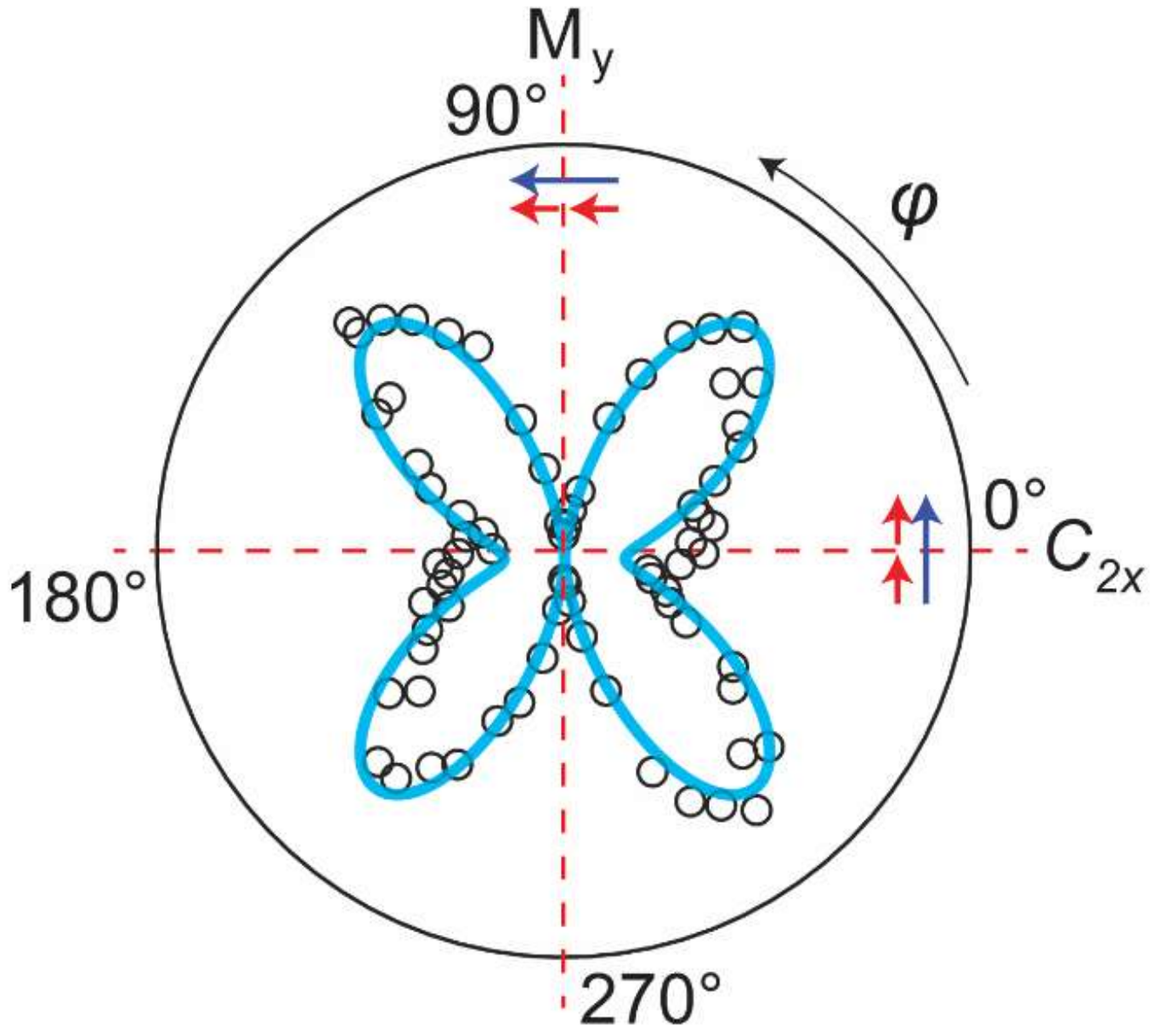

**Extended Data Fig. 5 | Identification of the in-plane crystallographic axes by polarization-resolved second-harmonic generation.** Polar plots of the second-harmonic response used to determine the in-plane crystallographic axes of bilayer $CrI_3$. Dots are SHG data points and the blue trace is the fitted curve (see Methods). The red and blue arrows indicate the polarization of fundamental and SHG light, respectively. The red dashed lines show the mirror plane $M_y$, where the SHG intensity is fully suppressed, and in-plane two-fold rotation axis $C_{2x}$.

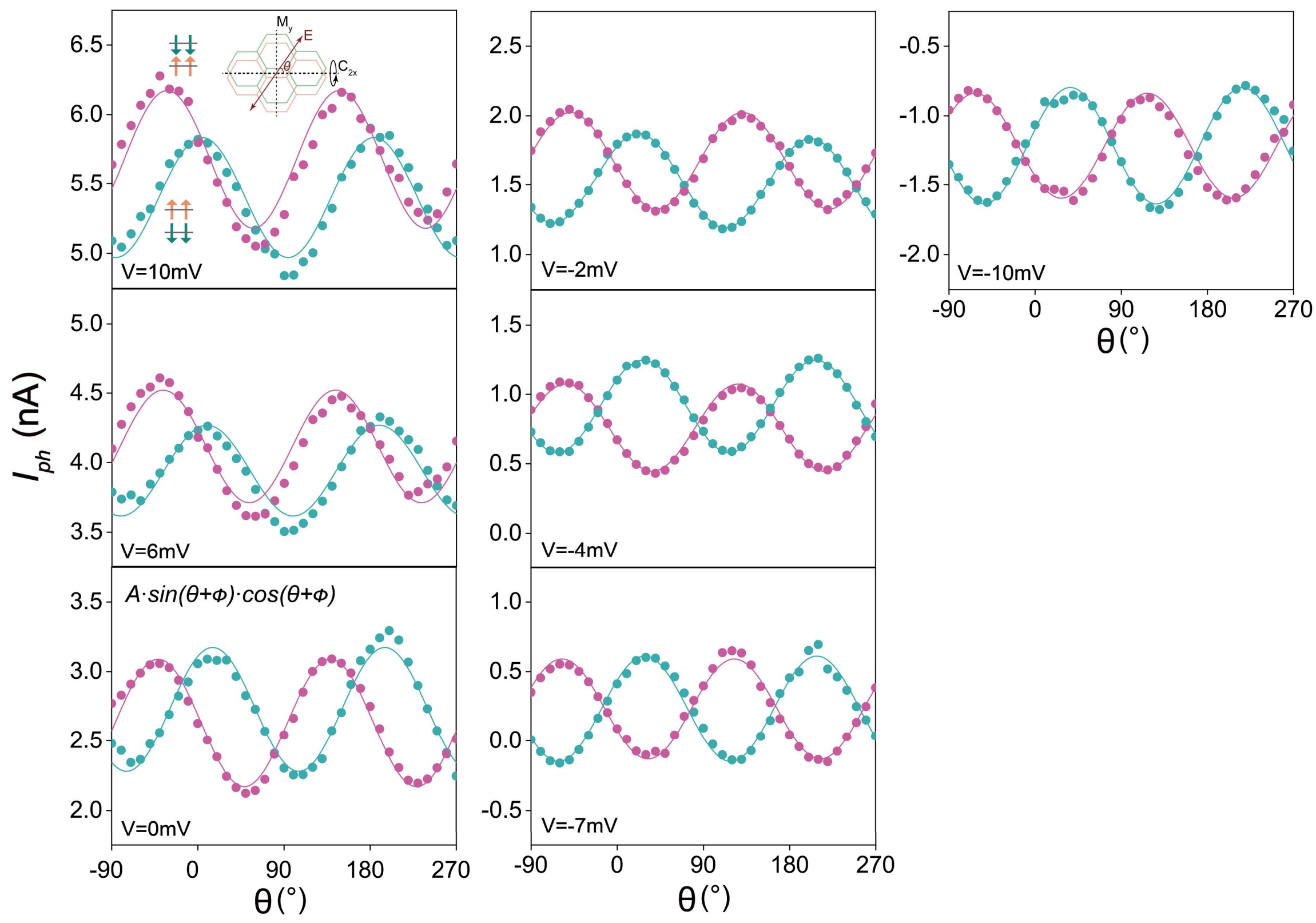

**Extended Data Fig. 6 | Linear polarization angle dependence of photocurrent for various junction biases.** Linear-polarization-angle-dependent photocurrent $I_{ph}(\theta)$ for the two AFM states of Device 3 under different voltage biases (10, 6, 0, -2, -4, -7, -10mV). The solid curves are fits to $I_{ph}(\theta) = Asin(\theta + \phi) \cdot cos(\theta + \phi)$. The oscillations retain the same periodicity, while phases $\phi$ of the two AFM states change with bias.